# GravitoMagneto-Hydrodynamics and Spacetime Turbulence in Early Universe[†]

Jiaxiang Liang[‡,1,2], Minghui Du[§,1], Peng Xu[*,1,2,3,4]

1. Center for Gravitational Wave Experiment, National Microgravity Laboratory, Institute of Mechanics, Chinese Academy of Sciences, Beijing 100190, China
2. Lanzhou Center of Theoretical Physics, Lanzhou University, Lanzhou 730000, China
3. Hangzhou Institute for Advanced Study, University of Chinese Academy of Sciences, Hangzhou 310024, China
4. Taiji Laboratory for Gravitational Wave Universe (Beijing/Hangzhou), University of Chinese Academy of Sciences, Beijing 100049, China

## Abstract

**Based on the gravitoelectromagnetic formalism and inspired by the rich analogies between electrodynamics and general relativity, we try one step further along this line and suggest a new counterpart in the gravitoelectromagnetic world analogue to the electromagnetic physics. A counterpart model of the MagnetoHydroDynamics that could help us to understand the possible new physics in tightly bounded spacetime-matter systems such as the case of extremely relativistic fluids in the early Universe. This new viewpoint also suggests a possible new form of spacetime-matter turbulence which may be tested through gravitational wave observations.**

---

† Essay written for the Gravity Research Foundation 2024 Awards for Essays on Gravitation.
‡ 220220939631@lzu.edu.cn
§ duminghui@imech.ac.cn
* xupeng@imech.ac.cn, corresponding author

# 1. Introduction

With Einstein's equation, and his vision of nature as geometry, space, time and matter are bound together in the so called geometrodynamics. "Matter tells spacetime how to curve, and spacetime tells matter how to move", as coined by J. Wheeler, gives to rich physics and deep implications. Remained still as the most fit theory confronted with the many stringent tests including gravitational wave detections, the highly nonlinearity in geometrodynamics predicted by general relativity (GR) plays important roles in the understandings of a wide range of physical processes, such as the studies of strong interactions in astrophysics, evolutions in the early stage of our Universe, and etc. Because of the complexity of the Einstein's equation, one generally need inspiring mathematical tools or theoretical methods. In this essay, we suggest that the Gravito-ElectroMagnetic (GEM) formalism of GR may help us to reveal and visualize a much stronger synergistic interaction between spacetime and matter for extremely relativistic fluids in early universe.

In GR, given a 3+1 foliation of spacetime and a congruence of observers with 4-velocity $u^\mu$ that orthogonal to the space slices. The Weyl curvature tensor $C_{\mu\nu\lambda\rho}$ can be split covariantly into two irreducible parts defined in space in terms of the projector $\gamma_{\mu\nu} = g_{\mu\nu} + u_\mu u_\nu$ [1].

$$E_{\mu\nu} = \gamma_\mu{}^m \gamma_\nu{}^n C_{mbnd} u^b u^d, \qquad B_{\mu\nu} = -\gamma_\mu{}^m \gamma_\nu{}^{n\,*} C_{mbnd} u^b u^d,$$

which mimic the separation of Maxwell tensor $F_{\mu\nu}$ into electric and magnetic fields. For weak field limit, in local Lorentz frames such analogue extends to the field equations satisfied by the gravito-electromagnetic fields [2,3]

$$\nabla \cdot \boldsymbol{E}_g = 4\pi G\rho + O(\boldsymbol{E}_g^2) + O(\boldsymbol{B}_g^2), \qquad \nabla \times \boldsymbol{B}_g = \frac{8\pi G}{c}\boldsymbol{J}_g + \frac{2}{c}\frac{\partial}{\partial t}\boldsymbol{E}_g + O(\boldsymbol{E}_g \times \boldsymbol{B}_g),$$

$$\nabla \times \boldsymbol{E}_g = -\frac{1}{2c}\frac{\partial}{\partial t}\boldsymbol{B}_g, \qquad \nabla \cdot \boldsymbol{B}_g = 0 + O(\boldsymbol{E}_g \cdot \boldsymbol{B}_g)$$

and the Lorentz-like force for test particles

$$\boldsymbol{F} = -m\boldsymbol{E}_g - 2m\frac{v}{c}\boldsymbol{B}_g.$$

Complicate nonlinear coupling terms appear when approaching high energy regime just like the case of nonlinear electrodynamics, and even for full GR such analogue retains still in some sense, see [1] for details.

Such analogue between GEM and electromagnetism had already been proved very useful, and had found applications in understanding the strong interactions between fluids and gravity in the studies of active galactic nucleus and their jets [3]. With the extended Navier-Stokes equation [3]

$$\rho\left[\frac{\partial \boldsymbol{u}}{\partial t} + \boldsymbol{u} \cdot \nabla \boldsymbol{u}\right] \cong \rho\left(\boldsymbol{E}_g + \boldsymbol{u} \times \frac{1}{c}\boldsymbol{B}_g\right) - \nabla p + \nabla \cdot (2\gamma\boldsymbol{\epsilon}),$$

the hydrodynamics of accretion disk driven also by the gravitomagnetic (GM) field could be studied. Here, $\gamma$ and $\boldsymbol{\epsilon}$ are the shear viscosity and tensor of the fluid, $\boldsymbol{u}$ the fluid velocity.

*In this essay, we try one step further along this line, and suggest a new counterpart in the GEM world analogue to the electromagnetic physics, that a counterpart model of the MagnetoHydroDynamics (MHD) that could help us to understand the possible new physics from such highly bounded spacetime-matter systems such as the case in early Universe.*

## 2. Gravito-MHD in Early Universe

MHD studies the dynamics of the interaction between magnetic fields and conducting fluids. It can be used to describe nonlinear dynamics, especially turbulences in such systems.

Currents are generated in conducting fluids when moving in a magnetic field and the Laplace force on such fluid is generated. According to Lenz's law, the Laplace force acts in a direction opposite to the movement of the fluid. Consequently, there are reciprocal interactions between the magnetic field and conducting fluid: the shapes of magnetic field lines are altered by the fluid, while the magnetic fields in turn exert a pulling force. The magnetic Reynolds number $R_m = \frac{Lu}{\eta}$ represents the strength of this interaction, where $L$ is the scale of fluid motion, $\eta = \frac{1}{4\pi\sigma}$ is the magnetic resistivity and $\sigma$ is the conductivity. when $R_m \to \infty$, the system is referred to as an ideal MHD. For an ideal MHD, the magnetic flux of any surface is always conserved. In other words, the magnetic field lines are frozen into the MHD fluid. Magnetic field lines and fluids are completely coupled and move together.

Therefore, according to the previous discussions, one key question that needs to be answered is that whether the Gravito-MagnetoHydroDynamics (GMHD) has a large gravitomagnetic Reynolds number $R_{gm}$ to be regarded as ideal, and whether the Alfvens theory still holds for GMHD?

Now, we focus on the relativistic fluids in early stage of our Universe, such as the epoch of the first-order electroweak phase transition, it is possible to approach such extreme condition with large $R_{gm}$. During this stage, phase bubbles were generated and collides between them took place, releasing a large amount of energy and building up turbulence cascades in the cosmic plasma. The Reynolds number of the fluid $R \sim 10^{16}$ and $R_m \sim 10^{24}$[5][6].

For the GM Reynolds number $R_{gm} = \frac{Lu}{\eta_g}, \eta_g = \frac{1}{4\pi\sigma_g}$. We can assume the wall speed $v_w$ of phase bubbles is comparable with the fluid velocity $u$. The Subsonic deflagrations,

Detonations and Supersonic deflagrations (hybrids) models[5] indicate that $v_w \approx 0.87c$ (detonation) or $v_w \approx c_s$ (deflagration)[5][6], where $c_s = 1/\sqrt{3}\,c$ is the sound speed in the relativistic fluid. As for the scale of fluid motion $L$, it is natural to set the diameter of phase bubbles as the featured size of $L = 2v_w/\beta$, with $\beta^{-1}$ is the characteristic time scale for the phase transition. The turbulence stirring time is equivalent to the duration of the phase transition. The turbulent cascade will stops when the dissipative effect caused by viscosity becomes important, and this happens at scales smaller than the Kolmogorov microscale $\lambda$, defined by

$$Re(\lambda) = \frac{u_\lambda}{\nu}\lambda \equiv 1, \qquad \text{hence} \quad \frac{L}{\lambda} = \frac{u_\lambda}{u_L}Re(\lambda) = Re(\lambda)^{\frac{3}{4}}.$$

$u_\lambda$ and $u_L$ are the velocities of turbulence at scales $\lambda$ and L respectively, and $\nu$ the kinematic viscosity. The initial scales of the turbulent inertial range can be determined by the ratio $L > 10^{10}\lambda$ [6]. For the effective "conductivity" $\sigma_g$ in GMHD, defined as $\boldsymbol{J} = \sigma_g \boldsymbol{E_g}$, considering the continuity equation we have

$$\frac{\partial \rho}{\partial t} \approx - \nabla \cdot (\sigma_g \boldsymbol{E_g})$$

Considering the isotropic and uniform conditions on cosmic scales, it is natural to assume $\sigma_g$ depends only on time. Then, we have

$$\frac{\partial \rho}{\partial t} = - \sigma_g 4\pi G \rho,$$

In the radiation-dominated era, the energy densities are $\rho \propto a(t)^{-4}$, and $\sigma_g = \frac{H(t)}{\pi G}$, with $H(t) = \frac{a'(t)}{a(t)} = G^{\frac{1}{2}}\hbar^{\frac{3}{2}}K_B{}^2 T^2$ the Hubble parameter. Here, $\hbar$ denotes the Planck constant, $K_B$ the Boltzmann constant, and $T$ the temperature. For electroweak phase transition, $T \approx 10^{15}$K, therefore $\sigma_g \approx 10^{39}$ kg·s·m$^{-3}$, which is extremely large. To conclude, we have $R_{gm} \to \infty$ in early universe in radiation-dominated era, where matter field fluctuations is extremely violent, gravity dominates and dissipations from other mechanisms including electrodynamics may not be important, and The GMHD system will be very close to ideal.

## 3. Alfven Theory for GMHD and Spacetime Turbulence

From previous discussions, the Alfven theory in ideal MHD is expected to hold for the case of ideal GMHD, that the GM field bounded with the relativistic fluid may build up cascades and form turbulence motions.

For GMHD, matter currents move in GM fields will be subjected to forces similar to the Laplace force. At the same time, we have shown that relativistic fluids in the early Universe have very high effective conductivities. When the motions of such fluid are altered by imposed GM field, the total GM fields will be compensated by the induced GM fields. The form of such tightly bounding is the same as that of MHD. So, the fluid and GM field will oscillate in a nonlinear way and continuously. We call this oscillation gravitational Alfven waves, see Figure 1 for illustration.

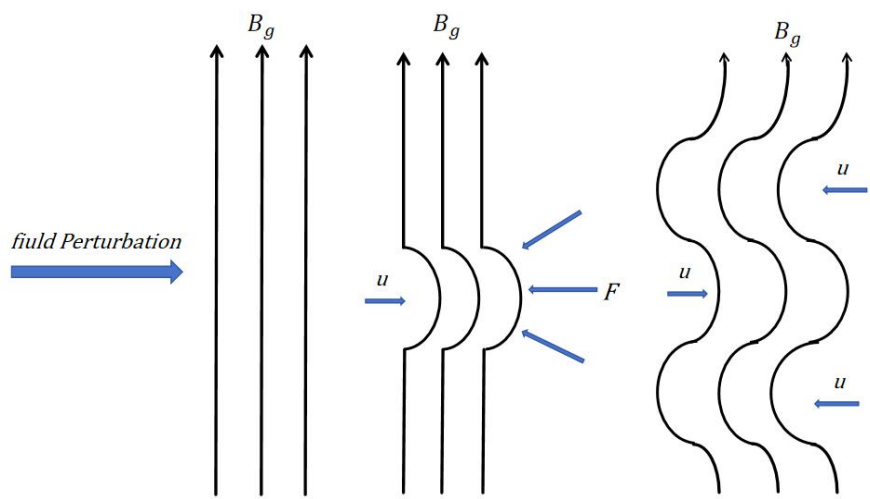

**Figure 1:** The oscillations caused by the disturbances give rise to the Alfven-like waves in GMHD.

Deeper perspectives may suggest that, with the collisions of Alfven wave-packets, energy cascades may be built up in GMHD, and energy is transferred from large scale to small scale and finally converted into heat in the dissipative scale. *This process occurs in both fluid and gravitomagnetic fields, and therefore the geometrodynamics of spacetime itself could turn into the form of turbulence,* which agrees with the recent foundlings summarized in [7].

At last, but not least, such new forms of turbulences of GMHD in early Universe may leave characteristic fingerprints in the stochastic gravitational wave background and could be traced in the precisely measured spectrum of the primordial GW backgrounds.